\documentclass[aps,prl,twocolumn,superscriptaddress,amsmath,amssymb]{revtex4-2}
\usepackage{times}
\usepackage{mathptmx}
\usepackage{bm}
\usepackage{amssymb}

\usepackage{graphicx}
\usepackage{dcolumn}
\usepackage{bm}
\usepackage{xspace}
\usepackage{amsmath}
\usepackage{amsfonts}
\usepackage{flushend}
\graphicspath{{NAQIfigures/} }
\usepackage[colorlinks=true, linkcolor=blue, citecolor=blue, urlcolor=blue]{hyperref}
\usepackage{placeins} 

\usepackage{gensymb}
\usepackage{booktabs} 
\usepackage{multirow} 
\usepackage{array} 
\usepackage{adjustbox} 
\usepackage{lipsum} 

\usepackage{textcomp} 
\usepackage{url}
   
\usepackage{xurl}

\usepackage{xcolor}
\usepackage{etoolbox}
\makeatletter
\newif\ifcolorrefs
\colorrefstrue
\newif\ifmatched

\begin{document}
	\title{Experimental certification of the Nonlocal Advantage of Quantum Imaginarity}
	\author{Jian-Hao Wu} 
	\affiliation{College of Physics and Optoelectronic Engineering, \href{https://ror.org/04rdtx186}{Ocean University of China}, Qingdao 266100, China.}
    \author{Kai-Yu Yuan} 
   	\affiliation{ China Mobile (Suzhou) Software Technology Company Limited, Suzhou 215163, China}
	\author{Yun-Xiang Tian} 
	\affiliation{College of Physics and Optoelectronic Engineering, \href{https://ror.org/04rdtx186}{Ocean University of China}, Qingdao 266100, China.}
	\author{Hao-Ran Tan} 
	\affiliation{College of Physics and Optoelectronic Engineering, \href{https://ror.org/04rdtx186}{Ocean University of China}, Qingdao 266100, China.}
	\author{Yan-Xin Rong}
	\affiliation{College of Physics and Optoelectronic Engineering, \href{https://ror.org/04rdtx186}{Ocean University of China}, Qingdao 266100, China.}
	\author{Zhen Shang} 
	\affiliation{College of Physics and Optoelectronic Engineering, \href{https://ror.org/04rdtx186}{Ocean University of China}, Qingdao 266100, China.}
	\author{Yong-Jian Gu} 
	\affiliation{College of Physics and Optoelectronic Engineering, \href{https://ror.org/04rdtx186}{Ocean University of China}, Qingdao 266100, China.}
	\affiliation{Engineering Research Center of Advanced Marine Physical Instruments and Equipment (Ministry of Education), \href{https://ror.org/04rdtx186}{Ocean University of China}, Qingdao 266100, China.}
	\affiliation{ Qingdao Key Laboratory of Advanced  Optoelectronics, \href{https://ror.org/04rdtx186}{Ocean University of China}, Qingdao 266100, China}
	\author{Ya Xiao}\email{xiaoya@ouc.edu.cn}
	\affiliation{College of Physics and Optoelectronic Engineering, \href{https://ror.org/04rdtx186}{Ocean University of China}, Qingdao 266100, China.}
	\affiliation{Engineering Research Center of Advanced Marine Physical Instruments and Equipment (Ministry of Education), \href{https://ror.org/04rdtx186}{Ocean University of China}, Qingdao 266100, China.}
	\affiliation{ Qingdao Key Laboratory of Advanced Optoelectronics, \href{https://ror.org/04rdtx186}{Ocean University of China}, Qingdao 266100, China}
	\date{\today}

\begin{abstract}
Quantum imaginarity is a distinct resource in quantum information theory, yet its nonlocal properties have not been fully explored. Here, we report an experimental study of the nonlocal advantage of quantum imaginarity (NAQI), in which local measurements on one subsystem can steer the average imaginarity of the conditional states of the other subsystem beyond the corresponding classical bound. The $l_1$-norm of imaginarity inequality is adopted as an experimentally accessible witness. A hybrid optimization algorithm that combines a genetic algorithm with sequential quadratic programming is developed to avoid local optima and accelerate the search for optimal measurement settings. Using polarization-encoded photonic qubits, we prepare two classes of two-qubit Bell-diagonal states and experimentally characterize the imaginarity of the conditional states following local measurements. Clear violations of the $l_1$-norm of imaginarity inequality are observed, providing an experimental certification of NAQI. We further investigate the relationship among NAQI, the nonlocal advantage of quantum coherence (NAQC), and Bell nonlocality based on their respective inequality criteria.  For the two-qubit Werner states considered in this work, the regions of states that violate the respective criteria satisfy $\mathcal{D}_{\rm NAQC}\subset\mathcal{D}_{\rm NAQI}\subset\mathcal{D}_{\rm BN}$.  Our work provides an experimental realization of NAQI and a comparison of different forms of quantum nonclassicality in a photonic platform.
\end{abstract}.
	
\maketitle

\section{I. INTRODUCTION}    
Imaginary numbers play an important role in accurately describing the behavior of quantum systems~\cite{PhysRevLett.133.190201,P1,P2,P3}, yet their physical necessity has long been debated. This debate is motivated in part by the existence of alternative formulations and representations of quantum theory that can reproduce substantial aspects of its predictions. In particular, generalized probabilistic theories provide a broader framework for exploring alternative mathematical structures underlying quantum theory~\cite{Hardy2001,Barrett2007,Plavala2023}, while the Wigner function offers an alternative phase-space representation in which quantum states are described through quasi-probability distributions ~\cite{Wigner1932}. Real Hilbert-space quantum mechanics can reproduce many predictions of complex quantum  theory~\cite{Aleksandrova2013,McKague2009,Pal2008,Stueckelberg1960}
and even supports universal quantum computation~\cite{Rudolph_arxiv,Aharonov_arxiv}. These developments have therefore raised the question of whether imaginary numbers are operationally indispensable to quantum theory.

This view changed with the resource-theoretic framework for quantum imaginarity introduced by Hickey and Gour ~\cite{Gour2017,basis-dependent}, which treats the imaginary part of a quantum state as a distinct resource. Wu \textit{et al.} subsequently provided an operational demonstration of quantum imaginarity as a resource in optical setups for local state discrimination tasks ~\cite{discriminationtask}.  Since then, quantum imaginarity has attracted considerable attention. A variety of imaginarity measures have been proposed, including those based on the trace norm~\cite{basis-dependent}, robustness~\cite{16}, geometric imaginarity~\cite{15}, relative entropy and weight~\cite{17}, convex roof constructions~\cite{18}, optimal transport cost~\cite{19}, $\epsilon$-measure~\cite{20}, Jensen-Shannon divergence~\cite{21}, and entropic measures~\cite{22,23,24,25}.  Diverse applications of quantum imaginarity have also emerged in fields such as information hiding~\cite{informationhiding,26}, multiparameter metrology~\cite{27}, machine learning~\cite{28}, pseudorandomness~\cite{29}, outcome statistics of linear-optical experiments~\cite{30}, Kirkwood-Dirac quasiprobability distributions~\cite{31,32,33,34}, weak-value theory~\cite{35}, broadcasting~\cite{41,42} and channel discrimination~\cite{15}. The quantum imaginarity of Gaussian states \cite{Gaussianstates,36} and of quantum channels~\cite{37} has also been investigated. 

The connections between imaginarity, holism, and nonlocality have also been an active line of research. Whether Bell nonlocality can be simulated with real-valued states alone has been examined in different scenarios~\cite{P1,P2,P3,basis-dependent}, related holistic properties have been investigated by local and bilocal tomography in generalized probabilistic theories~\cite{Hardy2011,Wootters2010,McKague2009,Erba_arXiv,Weilenmann_arXiv,Hoffreumon_arXiv,Hita_arXiv,Volovich_arXiv,Feng_arXiv,Song_arXiv,Ying_arXiv}, and the communication advantage of Wu \textit{et al.} also provides an operational connection between imaginarity and multipartite quantum correlations~\cite{discriminationtask}. More recently, Wei and Fei introduced NAQI~\cite{NAQI}, in which local measurements on one subsystem can steer an average imaginarity in the conditional state of the other qubits beyond a classical bound. Here ``nonlocal'' refers to a steering-like scenario and should be distinguished from Bell nonlocality, which is characterized by the failure of local hidden-variable models. 

NAQI is closely related to the nonlocal advantage of quantum coherence (NAQC) proposed by Mondal \textit{et al.}, where local measurements can steer the average coherence of the conditional states of the other subsystem beyond the corresponding classical bound~\cite{NAQC,Ding2019}.  Unlike imaginarity and coherence defined on single-party systems~\cite{Streltsov2017}, NAQI and NAQC are formulated in operational steering-like scenarios, making them suitable for probing the interplay between local resources and multipartite correlations. In contrast, Bell nonlocality reflects the failure of local hidden-variable descriptions of the full correlation statistics~\cite{Brunner_Rev2014}, whereas NAQI and NAQC quantify the remote steering of imaginarity and coherence, respectively. Investigating the relationships among these three nonclassical correlations is therefore meaningful, as it clarifies how different quantum resources are revealed by different operational criteria and may provide insights into their potential applications. Although NAQC and Bell nonlocality have been compared for certain families of states~\cite{NAQC,NAQCandBell}, and NAQI has been studied in relation to entanglement and steering~\cite{NAQI,CorRelation1,Steeringdefine}, a systematic comparison of the three nonclassical correlations on the same experimental platform remains to be further explored.

In this work, we experimentally investigate NAQI and its relationship to NAQC and Bell nonlocality using polarization-encoded photonic qubits.   We adopt the $l_1$-norm of imaginarity as an experimentally accessible criterion and develop a hybrid optimization procedure combining a genetic algorithm with sequential quadratic programming (GA-SQP)~\cite{GA_new1,GA_new2,SQP_new1,SQP_new2} which circumvents local optima and accelerates the identification of optimal measurement settings.  
We first prepare families of binary Bell-state mixtures and two-qubit Werner states,  providing an experimental demonstration of NAQI in a discrete-variable photonic system. We then focus on two‑qubit Werner states and compare the regions of states that exhibit NAQI, NAQC, and Bell nonlocality according to their respective inequality criteria. For the Werner states investigated here, the region exhibiting NAQC is contained within that exhibiting NAQI, which in turn is contained within the region exhibiting Bell nonlocality. Our work provides an experimental verification of NAQI and offers insights into the relationships among different forms of nonclassical correlations within specific families of states and operational criteria.

\section{II. theoretical background}
 \begin{figure}[htbp] 
		\centering
		\includegraphics[width=1\linewidth]{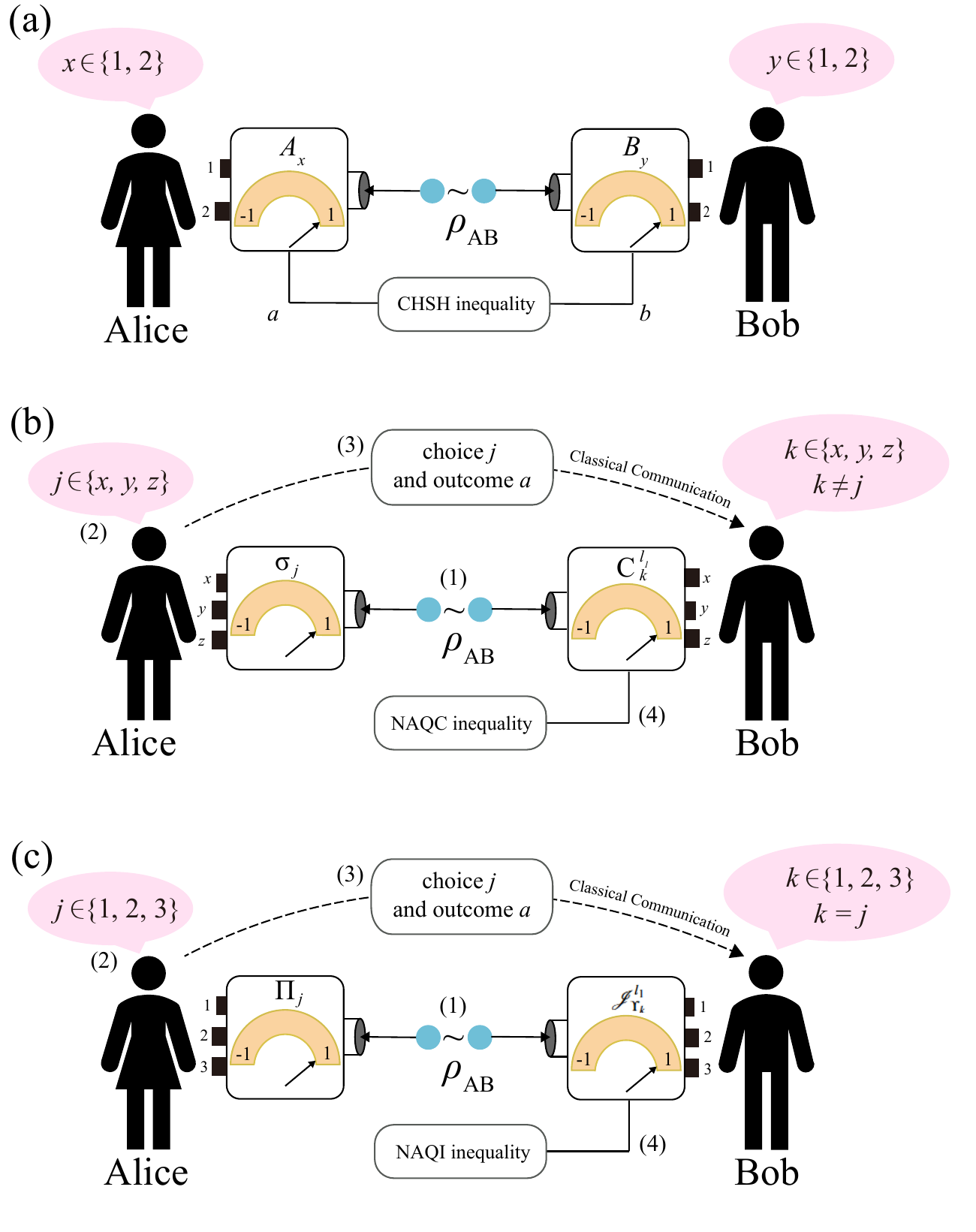} 
		\caption{Illustration of the Bell nonlocality, NAQC, and NAQI scenarios. (a) Bell nonlocality scenario. Alice and Bob share a bipartite quantum state $\rho_{\rm AB}$. Without communication, Alice and Bob choose observables from $\{A_1, A_2\}$ and $\{B_1, B_2\}$, respectively. After all the measurements, they calculate the correlation function, Bell nonlocality is certified if and only if the Ineq.~(\ref{CHSH}) is violated. (b) NAQC scenario. The procedure consists of the following steps: (1) Alice and Bob share a bipartite quantum state $\rho_{\rm AB}$. (2) Alice measures one of three Pauli observables chosen from $\sigma_j$ ($j=x,y,z$). (3) Alice communicates her chosen setting $j$ and the corresponding outcome $a$ to Bob via a classical channel. (4) Based on this information, Bob measures the $l_1$-norm of coherence his conditional state in the $\sigma_k$ ($k\neq j$) basis. By repeating this procedure, NAQC is certified if and only if the Ineq.~(\ref{NAQC}) is violated. (c) NAQI scenario. The NAQI scenario is similar to the NAQC scenario, except that in the second step Alice performs one of three two-outcome projective measurements chosen from $\Pi = \{|\Pi_j^a\rangle\langle \Pi_j^a| \mid j = 1,2,3;\, a = \pm\}$, and in the fourth step, based on Alice's measurement information, Bob measures the $l_1$-norm of imaginarity of his conditional state in the $\{|\Upsilon_j^+\rangle, |\Upsilon_j^-\rangle\}$ basis, chosen from a set of MUBs $\Upsilon = \{|\Upsilon_k^b\rangle \mid k = 1,2,3;\, b = \pm\}$. By repeating this procedure, NAQI is certified if and only if the Ineq.~(\ref{NAQI}) is achieved.}
	   \label{scenario} 
	\end{figure}

In this work, we focus on two-qubit states $\rho_\text{AB}\in\mathcal{H}_A\otimes\mathcal{H}_B$, with $\dim\mathcal{H}_A=\dim\mathcal{H}_B=2$. Before introducing the NAQI scenario, we first recall the Bell nonlocality and NAQC scenarios.

\textit{Bell nonlocality scenario.---} As depicted in Fig.~\ref{scenario}(a), in the Bell nonlocality scenario a two-qubit state $\rho_\text{AB}$ is shared between two spatially separated parties, Alice and Bob. Each party independently chooses one of two dichotomic observables: Alice measures $A_x$ and Bob measures $B_y$, where $x,y\in\{1,2\}$. The corresponding outcomes are denoted by $a,b\in\{-1,+1\}$. The correlation function is defined as
$\langle A_xB_y\rangle=\sum_{a,b=\pm1}ab\,p(a,b|x,y)=\operatorname{Tr}[\rho_\text{AB}(A_x\otimes B_y)]$.
The CHSH inequality reads~\cite{CHSH1}
\begin{equation}
|\langle A_1B_1\rangle+\langle A_1B_2\rangle+\langle A_2B_1\rangle-\langle A_2B_2\rangle|\leq 2.
\label{CHSH}
\end{equation}
A violation of Ineq.~\eqref{CHSH} certifies the existence of Bell nonlocality. The optimal measurement settings maximizing the violation of Ineq.~\eqref{CHSH} are obtained by the optimization method described in Ref.~\cite{CHSHsolve}.

\textit{NAQC scenario.---} The NAQC scenario is a steering-like task in which Alice steers the quantum coherence of Bob's subsystem, as depicted in Fig.~\ref{scenario}(b). Alice and Bob initially share a bipartite quantum state $\rho_\text{AB}$. Alice performs a projective measurement chosen from the set of Pauli observables $\{\sigma_j=|\eta_j^+\rangle\langle\eta_j^+|-|\eta_j^-\rangle\langle\eta_j^-| \mid j\in\{x,y,z\}\}$, where $|\eta_j^a\rangle\langle\eta_j^a|$ denotes the projector corresponding to outcome $a\in\{-1,+1\}$ of Alice's measurement $\sigma_j$. Consequently, Bob's state collapses to the conditional state $\rho_{B|\eta_j^a}=\frac{\operatorname{Tr}_A[(|\eta_j^a\rangle\langle\eta_j^a|\otimes \mathbb{I})\rho_\text{AB}]}{p(a|j)}$
with probability $p(a|j)=\operatorname{Tr}[(|\eta_j^a\rangle\langle\eta_j^a|\otimes\mathbb{I})\rho_\text{AB}]$, where $\mathbb{I}$ is the $2\times2$ identity matrix. After each measurement round, Alice communicates her chosen setting $j$ and the corresponding outcome $a$ to Bob via a classical channel. Bob then measures the $l_1$-norm of coherence of his conditional state in the eigenbasis of $\sigma_k$ ($k\neq j$), which is given by 
$C_k^{l_1}(\rho_{B|\eta_j^a}) = \sqrt{n_p^2+n_q^2}$, where $k,p,q\in\{x,y,z\}$ are all distinct, $a\in\{+1,-1\}$, and $(n_x,n_y,n_z)$ is the Bloch vector of $\rho_{B|\eta_j^a}$. 
A state is said to exhibit NAQC when it violates the inequality~\cite{NAQC}
\begin{equation}
\frac{1}{2}\sum_{\substack{j,k=x,y,z\\ j\neq k}}\sum_{a=\pm}p(a|j)\,C_k^{l_1}(\rho_{B|\eta_j^a}) \leq \sqrt{6}.
\label{NAQC}
\end{equation}

\textit{NAQI scenario.---} The NAQI scenario is analogous to the NAQC one, as illustrated in Fig.~\ref{scenario}(c). The difference is that Alice performs a projective measurement chosen from a set $\Pi = \{|\Pi_j^a\rangle\langle \Pi_j^a| \mid j = 1,2,3;\ a = \pm\}$ consisting of three two-outcome projective measurements, where $|\Pi_j^a\rangle\langle \Pi_j^a|$ denotes the projector corresponding to outcome $a$ of the $j$-th measurement setting. Bob’s conditional state is then expressed as
$
\rho_{B|\Pi_j^a} = \frac{\operatorname{Tr}_A[(|\Pi_j^a\rangle\langle \Pi_j^a| \otimes \mathbb{I})\rho_\text{AB}]}{p(a|j)},
$
with $p(a|j)=\operatorname{Tr}[(|\Pi_j^a\rangle\langle \Pi_j^a| \otimes \mathbb{I})\rho_\text{AB}]$. After each measurement round, Bob evaluates the  $l_1$-norm of quantum imaginarity of his conditional state $\mathcal{J}_{\Upsilon_k}^{l_1}(\rho)$ with respect to a set of two-dimensional mutually unbiased bases (MUBs) $\Upsilon = \{|\Upsilon_k^b\rangle \mid k = 1,2,3;\ b = \pm\}$. Mathematically, $\mathcal{J}_{\Upsilon_k}^{l_1}(\rho) = 2\left|\operatorname{Im}\langle \Upsilon_k^+|\rho|\Upsilon_k^-\rangle\right| = |\langle \sigma_y \rangle_{\Upsilon_k}|$, which represents the absolute imaginary parts of the off-diagonal elements in the basis $\{|\Upsilon_k^+\rangle,|\Upsilon_k^-\rangle\}$. For a qubit, this quantity equals the absolute expectation value of the Pauli observable $\sigma_y$ in that basis.

Wei \textit{et al.} proposed two NAQI criteria based on the $l_1$-norm of imaginarity and the relative entropy of imaginarity, respectively~\cite{NAQI}. As the $l_1$-norm criterion is more effective in detecting NAQI, we adopt it in this work. The NAQI parameter based on the $l_1$-norm of imaginarity is defined as
\begin{equation}\label{l1_norm}
\mathcal{N}(\rho_\text{AB})=\max_{\Upsilon,\Pi}\sum_{j,k,a}p(a|j)\,\mathcal{J}^{l_1}_{\Upsilon_{k}}(\rho_{B|\Pi_j^a}).
\end{equation}
Since both Alice's and Bob's subsystems are qubits, the maximization is performed over Alice’s three two-outcome projective measurements and Bob’s complete set of qubit MUBs, under the explicit parametrization given in Sec.~III.
A state $\rho_\text{AB}$ exhibits NAQI when $\mathcal{N}(\rho_\text{AB})$ exceeds the bound $\sqrt{5}$ imposed by the $l_1$-norm of imaginarity complementarity relation~\cite{NAQI}, i.e.,
\begin{equation}
\mathcal{N}(\rho_\text{AB}) >\sqrt{5}.
\label{NAQI_criterion}
\end{equation}

	\section{III. OPTIMAL MEASUREMENT SETTINGS}
	The NAQI criterion defined in Eq.~(\ref{NAQI_criterion}) clearly depends on the measurement settings. To fully detect NAQI for a given state, it is necessary to optimize Alice’s measurement set $\Pi$ and Bob's MUBs $\Upsilon$. Following Ref.~\cite{NAQI}, we parameterize them via angular variables
		\begin{equation}
			\begin{split}
				\{|\Pi_j^\pm\rangle\} =  \left\{\cos\frac{\theta_{j+1}}{2}\; |0\rangle + e^{i\phi_{j+1}}\sin\frac{\theta_{j+1}}{2}\; |1\rangle\, \text{,} \right. \\ \left.\sin\frac{\theta_{j+1}}{2}\; |0\rangle - e^{i\phi_{j+1}}\cos\frac{\theta_{j+1}}{2}\; |1\rangle\right\}\text{,}
			\end{split}
			\label{pi1}
		\end{equation}
	
	\begin{equation}
		\begin{split}
			\{|\Upsilon_1^{\pm}\rangle\} = \left\{\cos\frac{\theta_1}{2}|0\rangle + e^{i\phi_1}\sin\frac{\theta_1}{2}|1\rangle\text{,}\right. \\
			\left.\cos\frac{\theta_1}{2}|0\rangle - e^{i\phi_1}\sin\frac{\theta_1}{2}|1\rangle\right\}\text{,}
		\end{split}
		\label{M1}
	\end{equation}
	
	\begin{equation}
		\{ |\Upsilon_{2}^{\pm}\rangle\} = \left\{ \frac{|\Upsilon_{1}^{+}\rangle \pm |\Upsilon_{1}^{-}\rangle}{\sqrt{2}} \right\}\text{,}
		\label{M2}
	\end{equation}

	\begin{equation}
		\{ |\Upsilon_{3}^{\pm}\rangle\} = \left\{ \frac{|\Upsilon_{1}^{+}\rangle \pm i |\Upsilon_{1}^{-}\rangle}{\sqrt{2}} \right\}\text{,}
		\label{M3}
	\end{equation}
	where $|0\rangle$ and $|1\rangle$ are the eigenvectors of the Pauli matrix $\sigma_z$. Here, $j=1,2,3$, and the parameters $\theta_{p}\in[0,\pi]$, $\phi_{p}\in[0,2\pi]$ ($p=1,2,3,4$) are independent. Collecting them into $\mathbf{x}=(\theta_1, \theta_2, \theta_3, \theta_4,\phi_1, \phi_2, \phi_3, \phi_4)^\mathsf{T}$, we express the NAQI parameter as a function $\mathcal{N}(\mathbf{x};\rho_{\text{AB}})$ to be maximized subject to bound constraints  
	\begin{equation}
		\begin{aligned}
			\max_{\mathbf{x}} \quad & \mathcal{N}(\mathbf{x};\rho_{\text{AB}}) \\
			\text{s.t.} \quad & \mathbf{x}_\text{L} \leq \mathbf{x} \leq \mathbf{x}_\text{U}
		\end{aligned}\text{,}
		\label{opt_problem}
	\end{equation}
	with lower bound $\mathbf{x}_{\mathrm{L}}=(0,0,0,0,0,0,0,0)$ and upper bound $\mathbf{x}_{\mathrm{U}}=(\pi,\pi,\pi,\pi,2\pi,2\pi,2\pi,2\pi)$.

	\begin{figure}[htbp]
		\centering
		\includegraphics[width=1\linewidth]{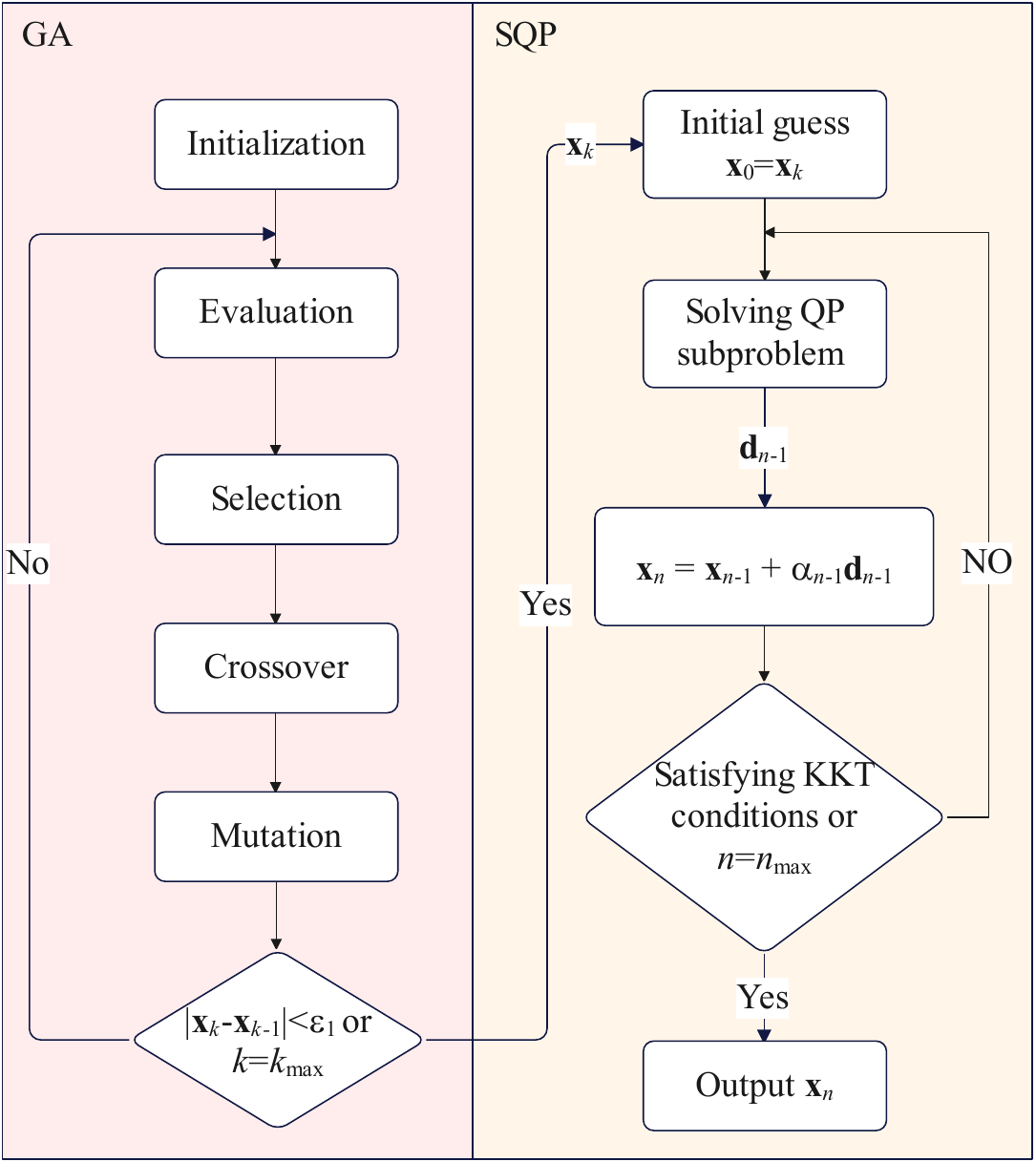} 
		\caption{Simplified flowchart of hybrid GA-SQP algorithm. 
			GA involves five procedures: initialization, evaluation, selection, crossover, and mutation. The last four steps repeat until either $|\mathbf{x}_k-\mathbf{x}_{k-1}|<\epsilon_1$ or $k=k_\text{max}$ is satisfied, where $\mathbf{x}_k$ and $\mathbf{x}_{k-1}$ are the optimal solutions at the $k$-th and $(k-1)$-th iteration, $k_\text{max}$ and $\epsilon_1$ are the predefined maximal number of iterations and the tolerance, respectively. Upon termination, the GA outputs the optimized solution $\mathbf{x}_k$ as the initial guess for SQP. In the $n$-th iteration of SQP, a constructed QP subproblem is solved to provide gradient information $\mathbf{d}_{n-1}$, and the solution is updated to $\mathbf{x}_{n} = \mathbf{x}_{n-1} + \alpha_{n-1} \mathbf{d}_{n-1}$, in which $\alpha_{n-1}$ is chosen to ensure sufficient decrease of the objective function. The SQP iteration terminates when either the KKT conditions are approximately satisfied or the maximum number of iterations $n_\text{max}$ is reached,  yielding the final optimal solution $\mathbf{x}_n$. }
		\label{fig:GA-SQP}
	\end{figure}

	To solve this nonlinear optimization problem, we develop a hybrid  GA-SQP algorithm. GA is an evolutionary optimization method based on the principles of natural selection~\cite{GA_new1,GA_new2}. It operates on a population of candidate solutions, each encoding a parameter vector $\mathbf{x}$. Since it requires no gradient information and can escape local optima, GA is well suited for global exploration. However, it typically suffers from slow convergence and limited precision. In contrast, SQP utilizes gradient information obtained by solving a sequence of quadratic programming (QP) subproblems, enabling rapid convergence to an optimal solution ~\cite{SQP_new1,SQP_new2}. Its primary drawback lies in its sensitivity to the initial guess, which, if poorly chosen, may lead to being trapped in a local optimum. Our hybrid GA-SQP optimization algorithm combines the global search capability of GA with the refinement efficiency of SQP. This hybrid approach allows efficient exploration of the parameter landscape while avoiding local maxima and ensuring high-precision convergence.
To further demonstrate the practical advantages of the hybrid GA-SQP optimization algorithm, we compare its performance with other algorithms in Appendix B.

	Fig.~\ref{fig:GA-SQP} shows a simplified flowchart of our algorithm. GA is first employed to generate a globally optimal initial candidate solution for a given quantum state $\rho_{\text{AB}}$. The algorithm begins by randomly initializing a population of $N$ candidate solutions, $\{\mathbf{x}_1^0, \mathbf{x}_2^0, \dots, \mathbf{x}_N^0\}$, within the feasible domain $\mathcal{D} = [0,\pi]^4 \times [0,2\pi]^4$, where $\mathbf{x}_i^0$ ($i=1, 2, \dots, N$) denotes $i$-th candidate solution at $0$-th iteration (i.e., initial settings), represented by an eight-dimensional parameter vector $\mathbf{x}_i^0 = (\theta_1^{0i}, \theta_2^{0i}, \theta_3^{0i}, \theta_4^{0i}, \phi_1^{0i}, \phi_2^{0i}, \phi_3^{0i}, \phi_4^{0i})^{\mathsf{T}}$.
	
	The objective function $\mathcal{N}(\mathbf{x};\rho_\text{AB})$ is then used to evaluate each candidate solution, and the $M$ ($M<N$) solutions corresponding to the largest values of $\mathcal{N}(\mathbf{x};\rho_\text{AB})$ are selected, while the best among them is denoted as $\mathbf{x}_0$; the remaining $N-M$ solutions are discarded. Retained solutions are randomly paired for crossover, generating $N-M$ offspring by exchanging components of their parameter vectors to further increase the value of $\mathcal{N}(\mathbf{x};\rho_\text{AB})$. To avoid premature convergence to local optima, a mutation operation is applied to each offspring by introducing small random perturbations to their parameters. This generates a new population of $N$ candidate solutions $\{\mathbf{x}_1^1, \mathbf{x}_2^1, \dots, \mathbf{x}_N^1\}$, where each candidate solution is again an eight-dimensional parameter vector $\mathbf{x}_i^1 = (\theta_1^{1i}, \theta_2^{1i}, \theta_3^{1i}, \theta_4^{1i}, \phi_1^{1i}, \phi_2^{1i}, \phi_3^{1i}, \phi_4^{1i})^{\mathsf{T}}$.
	
	The cycle of evaluation, selection, crossover and mutation is iteratively repeated. At the $k$-th iteration, a population of $N$ candidate solutions $\{\mathbf{x}_1^k, \mathbf{x}_2^k, \dots, \mathbf{x}_N^k\}$ is obtained, where each is given by $\mathbf{x}_i^k = (\theta_1^{ki}, \theta_2^{ki}, \theta_3^{ki}, \theta_4^{ki}, \phi_1^{ki}, \phi_2^{ki}, \phi_3^{ki}, \phi_4^{ki})^{\mathsf{T}}$. The best candidate solution in the current population is denoted as $\mathbf{x}_k$. The iteration terminates when either the change in the best solution falls below a predefined tolerance $\epsilon_1$ (i.e., $|\mathbf{x}_k - \mathbf{x}_{k-1}| < \epsilon_1$) or the maximum number of iterations $k_\text{max}$ is reached (i.e., $k=k_\text{max}$).

	Upon termination, GA outputs the optimized solution $\mathbf{x}_k$ as the initial guess for the SQP local optimization stage (i.e., $\mathbf{x}_0=\mathbf{x}_k$).
	For simplicity, we recast the original optimization problem in Eq.~(\ref{opt_problem}) into the standard minimization form
	\begin{equation}
		\begin{aligned}
			\min_{\mathbf{x}} \quad & f(\mathbf{x};\rho_\text{AB}) \\
			\text{s.t.} \quad & \mathbf{g}(\mathbf{x}) \leq \mathbf{0}
		\end{aligned}\text{,}
	\end{equation}
	where $f(\mathbf{x};\rho_\text{AB}) = -\mathcal{N}(\mathbf{x};\rho_\text{AB})$ and $\mathbf{g}(\mathbf{x}) = (\mathbf{x} - \mathbf{x}_U; \mathbf{x}_L - \mathbf{x})^\mathsf{T} \in \mathbb{R}^{16}$. 
	Let $\boldsymbol{\lambda}=(\lambda_1,\lambda_2,\dots,\lambda_{16})^\mathsf{T}$ denote the Lagrange multipliers and define the Lagrangian function $\text{L}(\mathbf{x}, \boldsymbol{\lambda}) = f(\mathbf{x}) + \boldsymbol{\lambda}^\mathsf{T} \mathbf{g}(\mathbf{x}).$
	The multipliers are typically initialized as $\boldsymbol{\lambda}^0$=(0,0,$\dots$,0)$^\mathsf{T}$ (i.e. at the 0-th iteration).

	\begin{figure*}[htbp]
		\centering
		\includegraphics[width=\textwidth]{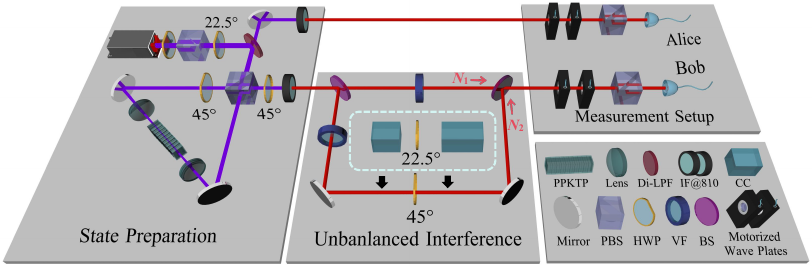}
		\caption{Experimental setup. (a) State Preparation: A 405 nm CW laser, whose polarization is set as $(|\text{H}\rangle+|\text{V}\rangle)/\sqrt{2}$ by a half-wave plate (HWP) at 22.5$^\circ$, pumps a type-II PPKTP crystal inside a polarization Sagnac interferometer, generating polarization-entangled photon pairs in $|\psi^+\rangle$. A HWP set at $45^\circ$ is then used to prepare the state $|\phi^+\rangle$. After filtering out the pump laser, one photon is sent to Alice, while  the other enters an unbalanced interferometer before being
        directed to Bob.
			(b) Unbalanced Interference: The short path leaves the state $|\phi^+\rangle$ unchanged, while the long path contains either a HWP at 45$^\circ$, which converts $|\phi^+\rangle$ into $|\psi^+\rangle$, or a combination of two sufficiently long calcite crystals (CCs) and a HWP at 22.5$^\circ$, which fully decoheres the polarization components, converting the state into the maximally mixed state $(\mathbb{I} \otimes \mathbb{I})/4$. The time delay between the short and long paths exceeds the photon coherence time. VFs in each path adjust the photon count ratio $N_1/N_2$, tuning the mixture parameter $p$ to prepare either the binary Bell-state mixtures $\rho_{\text{B}}(p)$ or the Werner states $\rho_{\text{W}}(p)$.
			(c) Measurement Setup: On both Alice's and Bob's sides, motorized QWPs, HWPs, and PBSs implement the required projective measurements. Photons are detected by single-photon detectors, and the signals are sent to coincidence counting for full state characterization and for evaluating the NAQI, NAQC, and Bell nonlocality violation parameters, $\Delta_{QI}$, $\Delta_{QC}$, and $\Delta_{BN}$.}
		\label{fig:setup} 
	\end{figure*}

	At the initial iteration (i.e. $1$-th iteration), a QP subproblem is constructed by taking a quadratic Taylor expansion of $\text{L}(\mathbf{x}, \boldsymbol{\lambda})$ around $\mathbf{x}_0$ while keeping the multipliers fixed at $\boldsymbol{\lambda}^0$,
	\begin{equation}
		\begin{aligned}
			\min_{\mathbf{d}} \quad & \frac{1}{2}\mathbf{d}^\mathsf{T} \nabla^2_{\mathbf{xx}} \text{L}(\mathbf{x}_0, \boldsymbol{\lambda}^0) \mathbf{d} + \nabla_\mathbf{x} \text{L}(\mathbf{x}_0,\boldsymbol{\lambda}^0)^\mathsf{T} \mathbf{d} \\
			\text{s.t.} \quad & \nabla \mathbf{g}(\mathbf{x}_0) \mathbf{d} + \mathbf{g}(\mathbf{x}_0) \leq \mathbf{0}
		\end{aligned}\text{,}
		\label{QP}
	\end{equation}
	with $\mathbf{d} = \mathbf{x} - \mathbf{x}_0$ and $\nabla_\mathbf{x} \text{L}(\mathbf{x}_0,\boldsymbol{\lambda}^0)=\nabla f(\mathbf{x}_0)+\nabla \mathbf{g}(\mathbf{x}_0)^\mathsf{T}\boldsymbol{\lambda}^0$. For our problem, the constraints are linear, therefore $\nabla \mathbf{g}(\mathbf{x})$ is a constant matrix
	\begin{equation}
		\nabla \mathbf{g}(\mathbf{x}) = 
		\begin{pmatrix}
			\mathbb{I}_8 \\
			-\mathbb{I}_8
		\end{pmatrix}
		\in \mathbb{R}^{16 \times 8} \text{,}
	\end{equation}
	and $\nabla^2_{\mathbf{xx}} \mathbf{g}(\mathbf{x})=\mathbf{0}$, where $\mathbb{I}_8$ is the $8 \times 8$ identity matrix. Consequently, $\nabla^2_{\mathbf{xx}} \text{L}(\mathbf{x}_0, \boldsymbol{\lambda}^0)$ reduces to $\nabla^2_{\mathbf{xx}} f(\mathbf{x}_0)$. Such a QP problem can then be solved systematically, producing a search direction $\mathbf{d}_0$ (i.e., the gradient information) and an updated multiplier vector $\boldsymbol{\lambda}^1=(\lambda_1^1,\lambda_2^1,\dots,\lambda_{16}^1)^\mathsf{T}$. A step size $\alpha_0 > 0$ is chosen to ensure a sufficient decrease of the objective function, and the primary solution is updated to $\mathbf{x}_{1} = \mathbf{x}_0 + \alpha_0 \mathbf{d}_0$.
	This process is repeated iteratively. At the \textit{n}-th ($n = 1,2,\dots,n_\text{max}$) iteration, we can construct a new QP subproblem analogous to Eq.~(\ref{QP}) by replacing $\mathbf{x}_0$ with $\mathbf{x}_{n-1}$ and $\boldsymbol{\lambda}^0$ with $\boldsymbol{\lambda}^{n-1}=(\lambda_1^{n-1},\lambda_2^{n-1},\dots,\lambda_{16}^{n-1})^\mathsf{T}$ obtained by solving the QP subproblem constructed in the $(n-1)$-th iteration. Here, $n_\text{max}$ is the predefined maximum number of iterations. By solving the newly constructed QP subproblem, the solution is updated to $\mathbf{x}_{n} = \mathbf{x}_{n-1} + \alpha_{n-1} \mathbf{d}_{n-1}$, and the Lagrange multipliers are updated to the solution $\boldsymbol{\lambda}^{n}$, where $\alpha_{n-1} > 0$ is determined again, and $\mathbf{d}_{n-1}$ is obtained from the $n$-th iteration. The iteration continues until either the Karush-Kuhn-Tucker (KKT) conditions~\cite{Karush1939, Kuhn1951, 2311.18707}  in Eq.~(\ref{KKT}) are approximately satisfied or the maximum number of iterations $n_\text{max}$ is reached. At a solution point $(\mathbf{x}_n,\boldsymbol{\lambda}^n)$, the KKT residuals are evaluated as 
		\begin{equation}
			\begin{aligned}
				\bigl\| \nabla f(\mathbf{x}_n) + \nabla \mathbf{g}(\mathbf{x}_n)^\mathsf{T} \boldsymbol{\lambda}^n \bigr\| &\leq \varepsilon_1, \quad \\
				\max_i |\lambda_i^n \, g_i(\mathbf{x}_n)| &\leq \varepsilon_2, 
			\end{aligned}
			\label{KKT}
		\end{equation}
		for $i=1,2,\dots,16$, while maintaining $\mathbf{g}(\mathbf{x}_n) \leq \mathbf{0}$ and $\boldsymbol{\lambda}^n \geq \mathbf{0}$, with the predefined tolerances $\varepsilon_1, \varepsilon_2 > 0$ predefined. Upon termination, the algorithm outputs the final solution $\mathbf{x}_n$, from which the maximum value $\mathcal{N}(\mathbf{x}_n;\rho_\text{AB})$ is obtained.

	\section{IV. EXPERIMENTAL SETUP AND RESULTS}
	We experimentally verify NAQI using two classes of Bell-diagonal states: the binary Bell-state mixtures
	\begin{equation}
		\rho_{\text{B}}(p) = p|\phi^+\rangle\langle\phi^+| + (1-p)|\psi^+\rangle\langle\psi^+|,
		\label{state1}
	\end{equation}
	and the two-qubit Werner states
	\begin{equation}
		\rho_{\text{W}}(p) = p|\phi^+\rangle\langle\phi^+| + \frac{1-p}{4}(\mathbb{I}\otimes\mathbb{I}),
		\label{state2}
	\end{equation}
	where mixture parameter $p \in [0,1]$ is a tunable parameter. $|\phi^+\rangle = \frac{1}{\sqrt{2}}(|\text{HH}\rangle + |\text{VV}\rangle)$ and $|\psi^+\rangle = \frac{1}{\sqrt{2}}(|\text{HV}\rangle + |\text{VH}\rangle)$, with $|\text{H}\rangle$ and $|\text{V}\rangle$ denoting horizontal and vertical polarizations, respectively. 
  Although both states in Eqs.~\eqref{state1} and \eqref{state2} are real-valued in the polarization computational basis $\{|\rm H\rangle,|\rm V\rangle\}$, they can still exhibit NAQI because quantum imaginarity is basis dependent. As discussed in Sec.~II, the $l_1$-norm of imaginarity  of Bob's conditional states $\mathcal{J}_{\Upsilon_k}^{l_1}(\rho)$ is evaluated with respect to the optimized MUBs $\{|\Upsilon_k^+\rangle,|\Upsilon_k^-\rangle\}$ rather than the $\{|\rm H\rangle,|\rm V\rangle\}$ basis. Consequently, a state that is real-valued in the polarization computational basis can have nonzero imaginary off-diagonal elements when expressed in the optimized MUBs. Experimentally, $\mathcal{J}_{\Upsilon_k}^{l_1}(\rho)$ is obtained from the absolute expectation value of $\sigma_y$ in the corresponding MUBs.

	Our experimental setup is shown in Fig.~\ref{fig:setup}. A 405~nm continuous-wave laser, prepared in $\frac{1}{\sqrt{2}}(|\text{H}\rangle+|\text{V}\rangle)$ using a half-wave plate (HWP) oriented at $22.5^\circ$, bidirectionally pumps a periodically poled $\mathrm{KTiOPO_4}$ (PPKTP) crystal in a polarization Sagnac interferometer, generating polarization-entangled photon pairs in the state $|\psi^+\rangle $ via type-II spontaneous parametric down-conversion (SPDC)~\cite{Experiment2}. A HWP set $45^\circ$, which exchanges the $|\mathrm{H}\rangle$ and $|\mathrm{V}\rangle$ polarizations, transforms this state into $|\phi^+\rangle$. After passing through interference filters (IFs) to remove the pump laser, one photon is sent directly to Alice, while the other is routed through an unbalanced interferometer before reaching Bob.

      \begin{figure}[htbp] 
		\centering
		\includegraphics[width=0.8\linewidth]{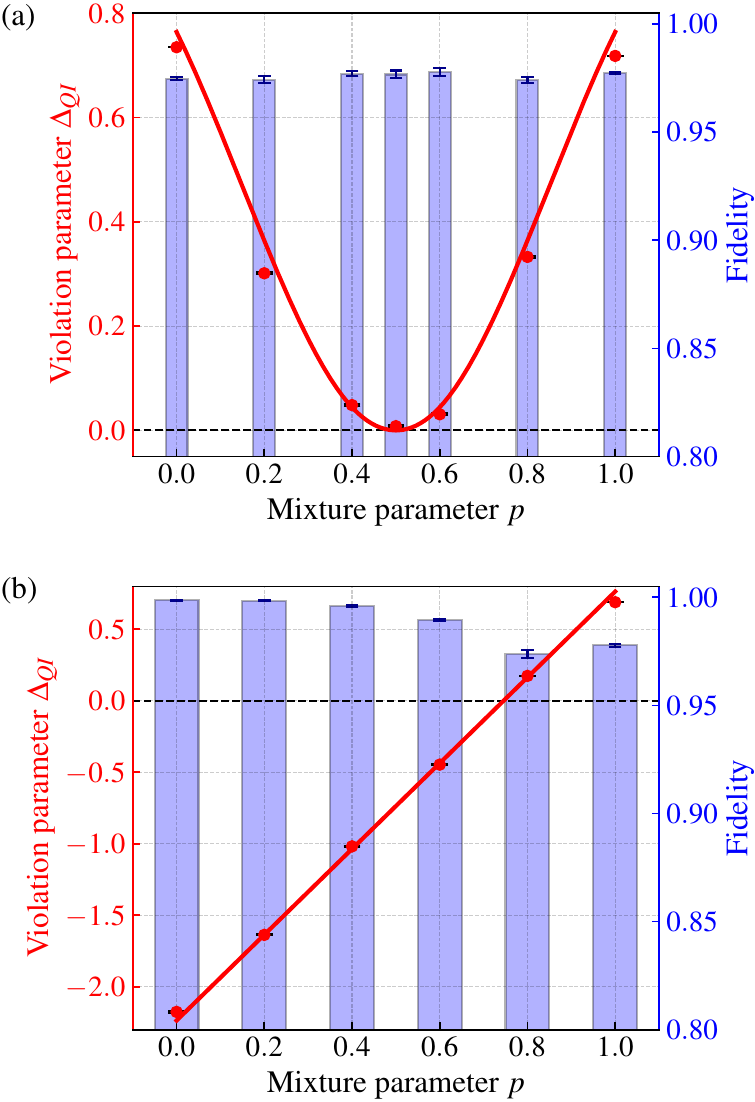} 
		\caption{Experimental results verifying NAQI. (a) Bell-state mixtures $\rho_{\text{B}}(p)$. (b) Werner states $\rho_{\text{W}}(p)$. In both panels, the solid light red curves show the theoretical predictions of violation parameter $\Delta_{QI}$ as a function of $p$, while red markers with error bars indicate experimental values. The blue bars represent the fidelity of the prepared states. Error bars, estimated from the Poisson statistics of two-photon coincidences, are on the order of $10^{-3}$.}
		\label{NAQI}
	\end{figure}
	\newcommand{\figsuba}{Fig.~\ref{fig:figure12}(a)\xspace}
	\newcommand{\figsubb}{Fig.~\ref{fig:figure12}(b)\xspace}

	The interferometer consists of two paths, one long and one short, with a time delay between them much greater than the coherence time of the SPDC photons \cite{Experiment1}. To prepare the binary Bell-state mixtures $\rho_{\mathrm{B}}(p)$, a HWP set at $45^\circ$ is inserted into the long path, transforming the polarization state in that path from $|\phi^+\rangle$ to $|\psi^+\rangle$. Since the state in the short path remains $|\phi^+\rangle$, incoherent recombination of the two paths at a beam splitter (BS) yields $\rho_{\mathrm{B}}(p)$. The mixing parameter $p$ is controlled by variable filters (VFs) placed in the two paths and determined from the measured photon counts $N_1$ and $N_2$ in the short and long paths, respectively, as  $p = N_1/(N_1 + N_2)$.

	To generate the two-qubit Werner states $\rho_{\text{W}}(p)$, a decohering module (inside the light dashed frame), consisting of two sufficiently long calcite crystals (CCs) and a HWP set at $22.5^\circ$, is inserted into the long path, while the original HWP set at $45^\circ$ is removed. The second CC is twice as long as the first, completely destroying coherence among the $|\text{HH}\rangle$, $|\text{HV}\rangle$, $|\text{VH}\rangle$, and $|\text{VV}\rangle$ components. 
	After passing through the CCs, the two-photon state $|\phi^+\rangle$ becomes the maximally mixed state $\frac{1}{4}(\mathbb{I}\otimes\mathbb{I})$. Incoherent recombination of the states from both paths then yields the Werner states $\rho_{\text{W}}(p)$, with $p$ again controlled by the VFs.
	
	A motorized quarter-wave plate (QWP), a motorized HWP, and a polarizing beam splitter (PBS) on both Alice's and Bob's sides allow implementation of the projective measurements required to investigate NAQI. The optimal measurement settings for Alice and Bob, obtained using our hybrid GA-SQP algorithm described in the previous section, are provided in Appendix A, and the corresponding codes can be found in Ref.~\cite{github}.

	We prepared seven binary Bell-state mixtures $\rho_{\text{B}}(p)$ with $p \in \{0.0,0.2,0.4,0.5,0.6,0.8,1.0\}$ and six two-qubit Werner states $\rho_{\text{W}}(p)$ with $p \in \{0.0,0.2,0.4,0.6,0.8,1.0\}$ to detect NAQI, achieving average fidelities of $0.9760 \pm 0.0013$ and $0.9890 \pm 0.0006$, respectively. Fig.~\ref{NAQI}(a) and (b) show the results for $\rho_{\text{B}}(p)$ and $\rho_{\text{W}}(p)$, with light red curves representing theoretical predictions of violation parameter $\Delta_{QI}=\max_{\Upsilon,\Pi}\sum_{j,k,a}p(a|j)\,\mathcal{J}^{l_1}_{\Upsilon_{k}}(\rho_{B|\Pi_j^a})-\sqrt{5}$ as a function of $p$, and red dots with error bars indicating experimental values. The region beyond the black line indicates the presence of NAQI. Consistent with theoretical expectations, $\rho_{\text{B}}(p)$ exhibits NAQI for all tested parameters except $p = 0.5$, whereas $\rho_{\text{W}}(p)$ exhibits NAQI for $p = 0.8$ and $1.0$. Small discrepancies between theory and experiment arise from nonideal state preparation and photon counting statistics. These results demonstrate that Alice can remotely steer the quantum imaginarity of Bob's subsystem via local projective measurements.
	
	\begin{figure}[htbp]
		\centering
		\includegraphics[width=0.8\linewidth]{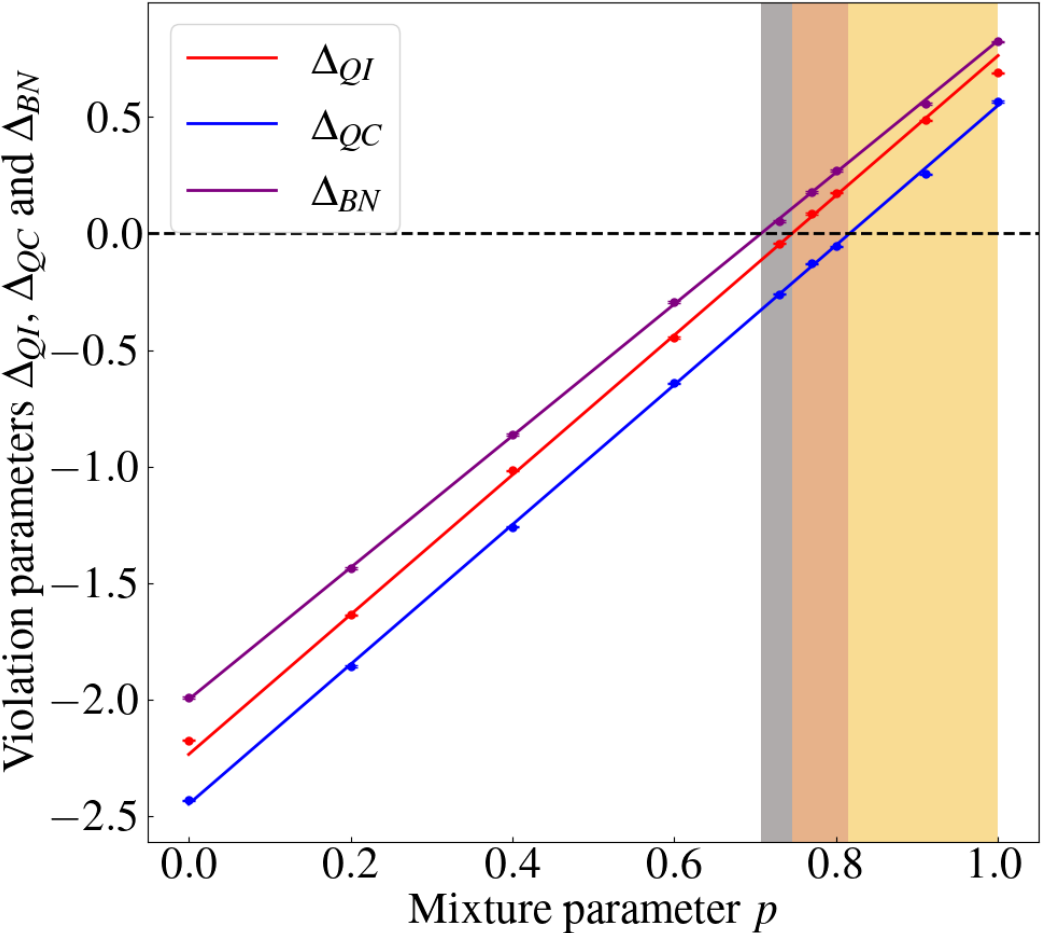} 
		\caption{Violation parameters $\Delta_{QI}$, $\Delta_{QC}$, and $\Delta_{BN}$ as a function of the mixture parameter $p$. Lines and markers show theoretical and experimental values. Light yellow, light red, and grey regions denote states exhibiting: Bell nonlocality only; Bell nonlocality and NAQI without NAQC; and all three correlations, respectively.
        }
		\label{hierarchical} 
	\end{figure}

	To elucidate the relationship among NAQI, NAQC, and Bell nonlocality,  we prepared nine two-qubit Werner states $\rho_{\text{W}}(p)$ with $p\in\{$ 0.00, 0.20, 0.40, 0.60, 0.73, 0.77, 0.80, 0.91, 1.00$\}$. The strengths of  NAQI, NAQC, and Bell nonlocality are quantified by the corresponding violation parameters $\Delta_{QI}$, $\Delta_{QC}=\frac{1}{2}\sum_{\substack{j,k=x,y,z\\ j\neq k}}\sum_{a=\pm}p(a|j)\,C_k^{l_1}(\rho_{B|\eta_j^a}) - \sqrt{6}$, and $\Delta_{BN}=|\langle A_1B_1\rangle+\langle A_1B_2\rangle+\langle A_2B_1\rangle-\langle A_2B_2\rangle|-2$. As shown in Fig. \ref{hierarchical}, all three violation parameters increase monotonically with $p$, with Bell nonlocality being detected first, followed by NAQI and then NAQC. Consequently, for the two-qubit Werner states and the specific criteria considered here, the regions of states that exhibit NAQC, NAQI, and Bell nonlocality satisfy $\mathcal{D}_{\rm NAQC}\subset\mathcal{D}_{\rm NAQI}\subset\mathcal{D}_{\rm BN}$. Furthermore, by tuning the mixture parameter, one can directly control the emergence of different nonclassical correlations, providing a practical means to engineer correlations for various quantum information protocols.

	\section{V. CONCLUSION}
In summary, we develop a hybrid GA-SQP algorithm to identify optimal measurement settings for detecting NAQI and experimentally certified NAQI in two families of Bell-diagonal states: the binary Bell-state mixtures and the two-qubit Werner states. Our results demonstrate that the average imaginarity of conditional states can be enhanced through local projective measurements, providing an experimental realization of the operational protocol for this recently introduced quantum resource. These results establish NAQI as a physically accessible and experimentally verifiable manifestation of quantum imaginarity in a bipartite photonic system.

Beyond its immediate application in NAQI detection, our hybrid GA–SQP algorithm can efficiently avoid local optima and accelerate the search for optimal measurement settings. This makes it broadly applicable to the certification of other nonclassical correlations in high‑dimensional or multipartite systems such as network nonlocality. Such correlations are essential for developing scalable quantum networks and enabling resource-efficient quantum information processing, where finding optimal measurement settings is often a central challenge.

Our results further reveal that, for the two-qubit Werner states and specific criteria considered here, the regions of states exhibiting NAQC, NAQI, and Bell nonlocality satisfy $\mathcal{D}_{\rm NAQC}\subset\mathcal{D}_{\rm NAQI}\subset\mathcal{D}_{\rm BN}$. This observation motivates further studies of these relations for more general state families and operational scenarios. More generally, the relationship between quantum imaginarity and other forms of nonclassicality, such as contextuality and quantum discord, remains an interesting direction for future investigation. Such studies may help clarify how different manifestations of quantum nonclassicality relate across different operational scenarios.

	\section*{acknowledgments}
	This work was supported by the National Key Research and Development Program of China (Grant No. 2025YFE0217700), the Shandong Provincial Natural Science Foundation (Grants No.  ZR2024LLZ003 and No. ZR2026LLZ016), the Fundamental Research Funds for theCentral Universities (Grant No. 202364008), and the YoungTalents Project at Ocean University of China (Grant No. 861901013107).

    \section*{DATA AVAILABILITY}
	The data that support the findings of this article are not publicly available. The data are available from the authors upon
	reasonable request.
    
	\quad	
	
	\appendix  
	\section*{{Appendix A: Optimal Measurement Settings for NAQI Certification}}
	\label{Optimal result}
	Table~\ref{optimal result1} lists the optimal measurement settings obtained using our hybrid GA-SQP algorithm for certifying NAQI in the binary Bell-state mixtures $\rho_{\text{B}}(p)$ defined in Eq.~(\ref{state1}), with 
	$p \in \{0.0, 0.2, 0.4, 0.5, 0.6, 0.8, 1.0\}$, while Table~\ref{optimal result2} provides the corresponding optimal measurement settings for the Werner states $\rho_{\text{W}}(p)$ defined in Eq.~(\ref{state2}), with $p \in \{0.0, 0.2, 0.4, 0.6, 0.8, 1.0\}$. The measurement setting parameters $\{\theta_1,\phi_1,\theta_2,\phi_2,\theta_3,\phi_3,\theta_4,\phi_4\}$ are defined in Eqs.~(\ref{pi1})--(\ref{M3}).
	
	\begin{table}[htbp]
		\centering
		\small
		\caption{Optimal settings for NAQI certification with  state $\rho_\text{B}(p)$.}
		\label{optimal result1}
		\begin{tabular*}{\columnwidth}{@{\extracolsep{\fill}}ccccccccc@{}}
			\toprule
			$p$ & $\theta_1/\degree$ & $\phi_1/\degree$ & $\theta_2/\degree$ & $\phi_2/\degree$ & $\theta_3/\degree$ & $\phi_3/\degree$ & $\theta_4/\degree$ & $\phi_4/\degree$ \\
			\midrule
			0.0 & 45.49 & 182.01 & 90.00 & 92.01 & 90.00 & 272.01 & 44.51 & 182.01 \\
			0.2 & 33.59 & 90.00 & 90.00 & 180.00 & 90.00 & 180.00 & 123.59 & 270.00 \\
			0.4 & 0.03 & 294.73 & 90.00 & 5.26 & 90.00 & 185.26 & 90.01 & 156.53 \\
			0.5 & 0.03 & 116.57 & 90.00 & 180.00 & 90.00 & 180.00 & 90.00 & 180.00 \\
			0.6 & 0.03 & 65.27 & 90.00 & 185.26 & 90.00 & 5.26 & 89.99 & 156.53 \\
			0.8 & 40.75 & 270.00 & 90.00 & 180.00 & 90.00 & 180.00 & 49.25 & 270.00 \\
			1.0 & 131.86 & 27.25 & 90.00 & 62.75 & 90.00 & 62.75 & 138.14 & 152.75 \\
			\bottomrule
		\end{tabular*}
	\end{table}
	
	\begin{table}[htbp]
		\centering
		\small
		\caption{Optimal settings for NAQI certification with  state $\rho_\text{W}(p)$}
		\label{optimal result2}
		\begin{tabular*}{\columnwidth}{@{\extracolsep{\fill}}ccccccccc@{}}
			\toprule
			$p$ & $\theta_1/\degree$ & $\phi_1/\degree$ & $\theta_2/\degree$ & $\phi_2/\degree$ & $\theta_3/\degree$ & $\phi_3/\degree$ & $\theta_4/\degree$ & $\phi_4/\degree$ \\
			\midrule
			0.0 & 40.67 & 61.45 & 40.98 & 156.85 & 56.00 & 332.42 & 77.44 & 66.53 \\
			0.2 & 134.29 & 131.03 & 90.00 & 138.97 & 90.00 & 138.97 & 44.29 & 228.97 \\
			0.4 & 121.79 & 307.18 & 90.00 & 142.82 & 90.00 & 322.82 & 148.21 & 232.82 \\
			0.6 & 43.47 & 309.07 & 90.00 & 140.93 & 90.00 & 320.93 & 46.53 & 230.93 \\
			0.8 & 130.61 & 27.60 & 90.00 & 62.40 & 90.00 & 242.40 & 139.39 & 152.40 \\
			1.0 & 55.05 & 31.59 & 90.00 & 58.41 & 90.00 & 58.41 & 34.95 & 148.41 \\
			\bottomrule
		\end{tabular*}
	\end{table}
	
\section*{Appendix B: Comparison of the hybrid GA-SQP optimization algorithm with other optimization methods}
\label{app}

To assess the performance of the hybrid GA-SQP algorithm, we compare it with Mathematica's \texttt{Maximize} and \texttt{NMaximize} solvers and the standard SQP algorithm for the optimization of the NAQI parameter  $\mathcal{N}(\rho_\text{AB})$ defined in Eq.~(\ref{l1_norm}). The comparison is carried out for the binary Bell-state mixtures $\rho_{\rm AB}(p)$ given in Eq.~(\ref{state1}), with $p$ varied from 0 to 1 in steps of 0.1. For each $\rho_{\rm AB}(p)$, the standard SQP algorithm is initialized at 50 randomly selected points. For the GA-SQP algorithm, the genetic algorithm is run five times to generate five candidate initial points, with a population size of $N=30$ and 20 iterations per run; these candidate points are subsequently refined by standard SQP algorithm. To prevent excessively long runtimes, a maximum wall-clock time of 120s is imposed for each optimization run.

\begin{table}[htbp]
	\centering
	\small
	\caption{Results obtained with different optimization algorithms. Here, $\delta_{12} = \mathcal{N}_{\rm GA-SQP}(\rho_{\rm AB}(p))-\mathcal{N}_{\rm SQP}(\rho_{\rm AB}(p))$, $\delta_{13} = \mathcal{N}_{\rm GA-SQP}(\rho_{\rm AB}(p))-\mathcal{N}_{\rm NMax}(\rho_{\rm AB}(p))$, and $\delta_{23} = \mathcal{N}_{\rm SQP}(\rho_{\rm AB}(p))-\mathcal{N}_{\rm NMax}(\rho_{\rm AB}(p))$, where $\mathcal{N}_{\rm GA-SQP}(\rho_{\rm AB}(p))$, $\mathcal{N}_{\rm SQP}(\rho_{\rm AB}(p))$, and $\mathcal{N}_{\rm NMax}(\rho_{\rm AB}(p))$ denote the optimized values obtained with the GA-SQP algorithm, the standard SQP algorithm, and the \texttt{NMaximize} solver, respectively. The corresponding runtimes are denoted by $t_1$, $t_2$, and $t_3$, respectively, with s denoting seconds.}
	\label{tab:comparison}
	\begin{tabular*}{\columnwidth}{@{\extracolsep{\fill}}ccccccc@{}}
		\toprule
		$p$ & $\delta_{12}/\times10^{-10}$ & $\delta_{13}/\times10^{-3}$ & $\delta_{23}/\times10^{-3}$ & $t_{1}/s$ & $t_{2}/s$ & $t_{3}/s$ \\
		\midrule
		0.0 & 0.00 & 0.00 & 0.00 & 3.17 & 3.55 & 28.74 \\
		0.1 & 0.00 & 0.00 & 0.00 & 1.18 & 2.31 & 31.63 \\
		0.2 & 0.00 & 0.00 & 0.00 & 1.23 & 2.22 & 24.70 \\
		0.3 & 0.00 & 8.32 & 8.32 & 1.28 & 2.51 & 21.40 \\
		0.4 & 0.00 & 0.00 & 0.00 & 1.24 & 2.41 & 21.17 \\
		0.5 & 0.00 & 0.00 & 0.00 & 1.32 & 2.38 & 27.94 \\
		0.6 & 0.00 & 7.32 & 7.32 & 1.28 & 2.38 & 23.36 \\
		0.7 & 0.00 & 8.32 & 8.32 & 1.23 & 2.37 & 21.81 \\
		0.8 & 0.00 & 0.00 & 0.00 & 1.19 & 2.25 & 21.49 \\
		0.9 & 0.00 & 0.00 & 0.00 & 1.15 & 2.89 & 26.14 \\
		1.0 & 0.00 & 0.00 & 0.00 & 1.14 & 1.77 & 29.15 \\
		\bottomrule
	\end{tabular*}
\end{table}

All calculations are performed on the same Windows 11 laptop equipped with a 13th-generation Intel Core i5-1340P processor and 16 GB of RAM, using MATLAB R2022b and Mathematica 12.0. The \texttt{Maximize} solver exceeds the 120-s time limit at $p=0.0$, $0.5$, and $1.0$, and the corresponding results are therefore omitted. The remaining results are summarized in Table~\ref{tab:comparison}. The GA-SQP algorithm yields optimized values that equal those obtained by the standard SQP algorithm to within a numerical precision of $10^{-10}$, while requiring substantially less computation time. In contrast, \texttt{NMaximize} requires considerably longer computation times and, at $p=0.3$, $0.6$, and $0.7$, yields lower optimized values than those obtained by the GA-SQP and SQP algorithms. These results indicate that the GA-SQP algorithm provides the best overall performance in terms of optimized values and computational time. This advantage may become more pronounced as the number of qubits and the Hilbert-space dimension increase. Small variations in wall-clock time may arise from fluctuations in computational load and software environment between different runs, whereas the optimized values remain stable within numerical precision. The corresponding source code is available in Ref.~\cite{github}.

	\bibliography{NAQIRefnew}
	
\end{document}